\documentclass[a4paper,11pt]{article}
\usepackage{jheppub,graphicx,floatflt}
\usepackage{amsmath,latexsym,amssymb,slashed}
\usepackage{enumerate}
\usepackage{mathrsfs}
\usepackage[colorlinks=true,%
            linkcolor=blue,%
            citecolor=blue,%
            urlcolor=blue]{hyperref}

\usepackage{graphicx} 
\usepackage{amsfonts}
\usepackage{amsmath} 
\usepackage{amssymb} 
\usepackage{float}

\usepackage{tipa}
\usepackage{textcomp}
\usepackage[T1]{fontenc} 
\usepackage[utf8]{inputenc}

\usepackage{tabularx} 

\usepackage{psfrag} 

\usepackage{dsfont}
\usepackage{multirow}

\usepackage{tikz}
\usepackage[compat=1.1.0]{tikz-feynman}

\newcommand\beal{\begin{align}}

\newcommand{\eq}[1]{\begin{equation}#1\end{equation}}
\newcommand{\spl}[1]{\begin{split}#1\end{split}}

\newcommand{\beq}{\begin{equation}}
\newcommand{\eeq}{\end{equation}}
\def\bea#1\eea{\begin{align}#1\end{align}}
\def\beal#1\eeal{\begin{subequations}\begin{align}#1\end{align}\end{subequations}}

\renewcommand{\i}{\ensuremath{\textnormal{i}}}

\newcommand{\boxedeq}[1]{
\begin{equation}
\fbox{
\rule[0.7cm]{0pt}{0pt}
$#1$
\rule[-0.45cm]{0pt}{0pt}
}
\end{equation}
}

\def\d{\text{d}}

\def\slashchar#1{\setbox0=\hbox{$#1$}           
\dimen0=\wd0                                 
\setbox1=\hbox{/} \dimen1=\wd1               
\ifdim\dimen0>\dimen1                        
\rlap{\hbox to \dimen0{\hfil/\hfil}}      
#1                                        
\else                                        
\rlap{\hbox to \dimen1{\hfil$#1$\hfil}}   
/                                         
\fi}

\def\del {\partial}

\renewcommand{\i}{\ensuremath{\textnormal{i}}}

\def\f {{\rm \texttt{f}}}
\def\mmm {\mathcal{M}}

\usepackage{xcolor}
\usepackage{color}

\newcommand{\cnote}[1]{}

\title{Dirac operator spectrum on a nilmanifold}

\author{Aldo Deandrea,}
\author{Fabio Dogliotti}
\author{and Dimitrios Tsimpis}

\affiliation{
Institut de Physique des Deux Infinis de Lyon \\
Universit\'{e} de Lyon, UCBL, UMR 5822, CNRS/IN2P3 \\
4 rue Enrico Fermi, 69622 Villeurbanne Cedex, France}

\emailAdd{deandrea@ipnl.in2p3.fr}
\emailAdd{dogliotti@ipnl.in2p3.fr}
\emailAdd{tsimpis@ipnl.in2p3.fr}

\abstract{
We obtain the spectrum of the Dirac operator on the three-dimensional Heisenberg nilmanifold $\mmm_3$, 
and its complete dependence on the metric moduli. As an application, we construct the four-dimensional low-energy 
effective action obtained by compactification of a seven-dimensional gauge-fermion theory on $\mmm_3$.
}

\begin{document}
\maketitle
\flushbottom
\setcounter{footnote}{0}
\renewcommand{\thefootnote}{\arabic{footnote}}
\setcounter{section}{0}

\section{Introduction and summary}

Compact spaces of negative scalar curvature have been considered in the context of extra-dimensional models in the past,  because of their extremely interesting properties for realistic model building, nevertheless they remain much less studied than their positively-curved counterparts. 
The present paper focuses in particular on the three-dimensional Heisenberg nilmanifold $\mmm_3$.  Nilmanifolds  are group manifolds based on nilpotent Lie algebras,  see {\it e.g.} \cite{2009arXiv0903.2926B} for a review. From the physics point of view, see  
 \cite{Kachru:2002sk, Grana:2006kf, Caviezel:2008ik,Camara:2009xy,Andriot:2015sia} 
for previous works,  they 
present an ideal playground for compactification and Kaluza-Klein (KK) reduction, as they are arguably the simplest non-trivial examples of negatively-curved manifolds on which exact calculations are possible. 

Obtaining the  four-dimensional low-energy effective theory from KK reduction of a higher-dimensional one, requires knowledge of the  spectrum of certain differential operators on the internal manifold, with the eigenmodes and eigenvalues corresponding to the fields and masses of the four-dimensional  theory.  
 In \cite{Andriot:2016rdd,Andriot:2018tmb} we studied the scalar and one-form spectrum on  $\mmm_3$. Knowledge of this part of the spectrum already allows for certain phenomenologically interesting applications to dark matter \cite{Andriot:2016rdd} and gauge-Higgs models \cite{Andriot:2020ola,Deandrea:2022aac}. However these works did not include the study of the fermion spectrum, whose knowledge is indispensable for realistic model building. In the present paper we fill this gap by studying the spectrum of the Dirac operator on $\mmm_3$.

Harmonic analysis on nilmanifolds, and  $\mmm_3$ in particular, has been considered before in the mathematical literature \cite{thang,Gordon, Gornet1, Gornet2, Muller, diracmath}, however the results are not always presented in a way accessible to physicists. Moreover the spectrum is typically computed using the canonical metric (the analogue of a square three-torus) and thus misses the dependence of the spectrum on the metric moduli. However, this dependence is an important  piece of information for physical applications, as it affects the masses of the fields of the theory. Obtaining the  complete metric moduli dependence of the spectrum is the main result of our paper. 

Furthermore, as already discussed in \cite{Andriot:2018tmb}, the  spectrum admits a  low-energy truncation to massless and  light massive modes. As an application of our results,  we use this truncation  to construct the four-dimensional effective action obtained by compactification of a seven-dimensional gauge-fermion theory on $\mmm_3$. 

The present work is part of the program initiated in \cite{Andriot:2016rdd,Andriot:2018tmb,Andriot:2020ola,Deandrea:2022aac} to explore the phenomenology of nilmanifold compactifications. 
As argued in those papers, compacification on nilmanifolds, or more general group manifolds, present several attractive features, in particular in the context of gauge-Higgs unification.  Besides their purely mathematical interest, our results will allow to complete these models by taking into account the KK reduction of the  fermionic sector of the seven-dimensional theory, thus taking a major step towards a more realistic phenomenology.  
 
The outline of the paper is as follows. In \S\ref{sec:2.1} we review the scalar spectrum on the Heisenberg nilmanifold and its dependence on the metric moduli. In \S\ref{sec:2.2} we derive the spectrum of the Dirac operator in the simplified case,  where only the dependence of the metric on the radii is taken into account. The general dependence of the spectrum on all  metric moduli is obtained  in \S\ref{sec:2.3}. The four-dimensional effective action arising from  reduction of a seven-dimensional gauge-fermion theory is contained in \S\ref{sec:3}. Our spinor conventions and the spectrum of the light eigen one-forms on $\mmm_3$ are discussed in the appendices \S\ref{app:fermions} and \S\ref{sec:b} respectively.
 
\section{Dirac operator spectrum on the Heisenberg nilmanifold}

Before addressing the spectrum of the Dirac operator on $\mmm_3$ in  \S \ref{sec:2.2}  and  \S \ref{sec:2.3}, it will be useful to give a brief review of the most general metric and  the associated scalar Laplacian eigenbases in \S \ref{sec:2.1}.

\subsection{Review of the scalar spectrum on $\mmm_3$}\label{sec:2.1}

The three-dimensional nilmanifold $\mmm_3$  is built from the nilpotent Heisenberg algebra
\eq{\label{heis}[V_1,V_2]=- \f\, V_3~,~~~[V_1,V_3]=[V_2,V_3]=0~,}
with structure constant $\f=-f^3{}_{12}$. The Maurer--Cartan one-forms $e^{a=1,2,3}$, which are dual to the vectors $V_a$ above, satisfy
\eq{\label{mcheis}
\d e^3= \f\, e^1\wedge e^2~;~~~\d e^1=0~;~~~\d e^2=0
~.}
These vectors and one-forms furnish bases of the tangent and cotangent spaces of $\mmm_3$ respectively. 
Using coordinates $x^{m=1,2,3} \in [0,1]$,  and constant radii $r^{m=1,2,3}$, we parametrize, 
\eq{\label{a}e^1=r^1\d x^1~;~~~e^2=r^2\d x^2~;~~~e^3=r^3\left(\d x^3+N x^1\d x^2\right)~;~~~N=\frac{r^1r^2}{r^3}\f \in\mathbb{Z}^*~.}
The vielbein $e^a{}_m$ and its inverse  $e^m{}_a \equiv (e^{-1})^m{}_a$ are given by $e^a = e^a{}_m \d x^m, \ V_a = e^m{}_a \del_m$. 
We use letters from the beginning, the middle of the  latin alphabet for flat, curved indices respectively. Explicitly the vectors read
\eq{\label{explv} 
V_1=  \tfrac{\partial}{\partial X^1}~,~~~ V_2=\tfrac{\partial}{\partial X^2} - \f X^1 \tfrac{\partial}{\partial X^3}~;~~~ V_3=\tfrac{\partial}{\partial X^3}~,}
where we have defined $X^{m=1,2,3}=r^m x^m$. 
Subject  to the  discrete identifications 
\eq{\label{heisids}
x^1\sim x^1+n^1~;~~~x^2\sim x^2+n^2~;~~~x^3\sim x^3+n^3- n^1N x^2~;~~~n^1,n^2,n^3\in\{ 0,1\}~,
}
the manifold $\mmm_3$ is compact, and is topologically a twisted (for $\f\neq0$) circle fibration, with fiber parameterized by $x^3$,  over a two torus base parameterized by $x^{1,2}$. The one-forms $e^a$ are invariant under \eqref{heisids}, thus globally defined. The most general metric on $\mmm_3$ is given by  \cite{Andriot:2016rdd}
\eq{\label{expl}\d s^2=\big(e^1+a e^3\big)^2+\big(e^2+b e^3\big)^2+\big(e^3\big)^2~,~~~a,b\in\mathbb{R}~.}
As explained in \cite{Andriot:2016rdd}, the parameters $a$, $b$ are moduli related to complex deformations of $T^2\subset\mmm_3$. 
Let us also note that that $\sqrt{g}= r^1 r^2 r^3$, so that the volume is given by
\eq{\label{vol} V=\int\d^3 x\sqrt{g}= r^1 r^2 r^3~.}
The eigen-modes of the scalar Laplacian operator on $\mmm_3$ are given by two distinct sets of orthonormal eigenfunctions. 
The first set  has a non-trivial dependence on $x^3$ and is given by  \cite{Andriot:2016rdd}
\eq{\spl{\label{uultcap2}
U_{k,l,n}(x^1,x^2,x^3)&=  \sqrt{\frac{r^2}{|N|V}}\frac{1}{\sqrt{2^n n! \sqrt{\pi}}}\, e^{2\pi K \i(X^3+\f\, X^1X^2)} e^{2\pi L \i X^1}\sum_{m\in\mathbb{Z}} e^{2\pi K M\i X^1} \\
&  \times \exp\left[-\frac{\i\pi K b}{b^2+1} \bigg(X^2 + \frac{M}{\f} + \frac{L}{K\f} \bigg) \bigg(\f a \bigg(X^2 + \frac{M}{\f} + \frac{L}{K\f} \bigg) - 2 \bigg) \right] \\
&  \times \Phi^{\sigma}_n\bigg(X^2 +\frac{M}{\f} + \frac{L}{K\f} - \frac{a}{\f(a^2+b^2+1)}\bigg)
\; ;\\
& \qquad \qquad \qquad \qquad \qquad \qquad l=0,\dots, |k|-1\; ,~k\in\mathbb{Z}^*\; ,~n\in\mathbb{N} \; ,
}}
where
\eq{\label{la}
i=1,2,3~,~~~ K=\frac{k}{r^3}~,~~L=\frac{l}{r^1}
~,~~ M=\frac{r^3}{r^1}m~;~~~m,k,l\in\mathbb{Z}\; ,}
and
\eq{\label{lambda}
\sigma=\frac{2\pi K }{b^2+1}(a^2+b^2+1)^{\frac12} \f
\; .}
The function $\Phi^{\sigma}_n$ is given by
\eq{\label{md}
\Phi^{\sigma}_n(z)=|\sigma|^{\frac14}\, \Phi_n(|\sigma|^{\frac12}z)\ , \quad \Phi_n(z)=e^{-\frac12 z^2}H_n(z)
\ ,\quad n\in\mathbb{N}\; ,}
where $H_n$ are the Hermite polynomials: $H_n(y)=(-1)^n e^{y^2} \del^n_y e^{-y^2}$. 
The Laplacian eigenvalues of the $U$-eigenfunctions are given by
\eq{\label{laplulcap}
\big(\nabla^2+ M^2_{k,l,n}\big)U_{k,l,n}=0~;~~~M^2_{k,l,n}=\frac{4\pi^2 k^2}{(r^3)^2(a^2+b^2+1)}\left[1+\frac{(2n+1)r^3|\f|}{2\pi |k |}(a^2+b^2+1)^{\frac32}\right]
\; .}
The second set of eigenfunctions of the scalar Laplacian on $\mmm_3$ have no dependence on the $x^3$ coordinate. They are given by
\eq{\label{52}V_{p,q}(x^1,x^2)= \frac{1}{\sqrt{V}}\, e^{ \i P X^1 } e^{ \i Q X^2 } \ , \quad P=\frac{2\pi p}{r^1},\ Q=\frac{2\pi q}{r^2},\ p,q \in \mathbb{Z}\ ,}
with eigenvalues
\eq{\label{53} (\nabla^2 + \mu_{p,q}^2)  V_{p,q}=0 \ , \quad \mu_{p,q}^2=  \frac{p^2}{(r^1)^2} + \frac{q^2}{(r^2)^2} +   \left(a \frac{p}{r^1} + b \frac{q}{r^2} \right)^2  \ .}

\subsection{Simple case: $a=b=0$}\label{sec:2.2}

As a warmup let us consider the Dirac operator eigenvalue problem on $\mathcal{M}_3$ in the simplified case $a=b=0$. 
The Dirac eigenvalue equation reads,
\bea\label{dep}
(D - \lambda) \psi = 0 ~;~~~ D\equiv\gamma^a e^{m}{}_a (\partial_{m} + \tfrac{1}{4}\omega_{m b c}\gamma^{bc} ) \;,
\eea
where   $\omega_{m b c}$ is the spin connection. The  independent  non-vanishing components of the flat spin connection 
$\omega_{a b c}=e^{m}{}_a \omega_{m b c}$ on $\mmm_3$  are given by 
\eq{
\omega_{123} =\omega_{231} = -\omega_{312} = \frac{\f}{2} \; .}
The Dirac operator then takes the form
\bea\label{dirac operator 1}
D=\gamma^aV_a  + i \frac{\f}{4}\; ,
\eea
where the $V_a$'s were given in \eqref{explv}. 
The $\gamma^a$'s can be taken to be the Pauli matrices, so that \eqref{dep} reduces to the following set of equations,
\eq{\spl{\label{depr}
V_3 \psi_1 + (V_1 -i V_2) \psi_2 - (\lambda - i\frac{\f}{4}) \psi_1 &= 0  \\
- V_3 \psi_2+( V_1 +i V_2) \psi_1  - (\lambda - i \frac{\f}{4} ) \psi_2 &= 0
\; .}}
Next we expand the two-component  Dirac spinor on the complete basis of Laplacian eigenfunctions on $\mmm_3$,\footnote{This expansion implicitly assumes that the Dirac spinor on $\mmm_3$ returns to its original value after a going around each of the three circles parameterized by $x^a$. This condition can be relaxed, leading to what is known as nontrivial spin structures  \cite{diracmath}.}
\bea
\label{psi_decomposition}
\psi
=
\begin{pmatrix}
 \psi_1 \\
\psi_2
\end{pmatrix}
= \sum_{k,l,n} 
\begin{pmatrix}
C^1_{k,l,n}   \\
C^2_{k,l,n} 
\end{pmatrix} u_{k,l,n} +
\sum_{p,q} 
\begin{pmatrix}
   D^1_{p,q} \\
 D^2_{p,q}
\end{pmatrix} v_{p,q}
 \; ,
\eea
where, following \cite{Andriot:2018tmb},  we use the notation $u_{k,l,n}$ and $v_{p,q}$, for the $a, b\rightarrow0$ limit of the functions given in \eqref{uultcap2} and \eqref{52} respectively. Since the two sets of functions are orthogonal to each other,  we can treat the two cases independently. 
The action of the $V_a$ on the $v$-basis reads, 
\eq{
V_1v_{p,q}=iP v_{p,q} ~;~~~ V_2v_{p,q}=iQ v_{p,q} ~, ~~~V_3v_{p,q}=0\; ,
} 
so that \eqref{depr} reduces to
\eq{\spl{
i(P -i Q) D_{p,q}^2 - (\lambda_{p,q} - i \frac{\f}{4}) D_{p,q}^1 &= 0  \\
i(P +iQ) D_{p,q}^1 - (\lambda_{p,q} - i \frac{\f}{4} ) D_{p,q}^2 &= 0  
\; .}}
From this system we deduce the eigenvalues 
\bea 
 -i \lambda_{p,q}= \frac{\f}{4} \pm \sqrt{P^2+Q^2}\; ,
\eea
and the corresponding eigen-spinors
\bea
\psi_{p,q}=C
\begin{pmatrix}
v_{p,q} \\
\alpha v_{p,q}
\end{pmatrix};
 \ \ \  \alpha = \pm\frac{\sqrt{P^2 + Q^2}}{(P-iQ)}\; ,
\eea
up to a normalization constant $C\in\mathbb{C}$. 

Let us now turn to the $u_{k,l,n}$ series. Using the action of the $V's$ on the basis, 
\eq{\spl{
\i V_3 u_{k,l,n} &= -\frac{ \vert \sigma \vert }{\f} sign(\sigma) u_{k,l,n} \\
V_2 u_{k,l,n} &= \frac{ \vert \sigma \vert^{\frac{1}{2}} }{2} \left(-\sqrt{2(n+1)} u_{k,l,n+1} + \sqrt{2n} u_{k,l,n-1} \right) \\
\i V_1 u_{k,l,n} &= -\frac{ \vert \sigma \vert^{\frac{1}{2}} }{2} sign(\sigma) \left( \sqrt{2(n+1)} u_{k,l,n+1} + \sqrt{2n} u_{k,l,n-1} \right)\; ,
}}
with $\sigma$ defined in \eqref{lambda}, we obtain 
\eq{
\spl{
\big[ \i\lambda_{k,l,n} + (\frac{\f}{4} + \frac{\sigma}{\f}) \big]C^1_{k,l,n} &= - \vert \sigma \vert^{\frac{1}{2}} \big[ p^+(\sigma) \sqrt{2n} C^2_{k,l,n-1} + p^-(\sigma) \sqrt{2(n+1)} C^2_{k,l,n+1} \big] 
\\
\big[\i\lambda_{k,l,n} + (\frac{\f}{4} - \frac{\sigma}{\f}) \big]C^2_{k,l,n} &= - \vert \sigma \vert^{\frac{1}{2}} \big[ p^-(\sigma) \sqrt{2n} C^1_{k,l,n-1} + p^+(\sigma) \sqrt{2(n+1)} C^1_{k,l,n+1} \big] 
\; ,
}}
where we have defined $p^{\pm}(\sigma) \equiv \frac{sign(\sigma) \pm 1}{2}$. We thus obtain
\bea\label{2.28}
 -\i \lambda_{k,l,n} =& \frac{f}{4} \pm \sqrt{ \left( \frac{\sigma}{f} \right)^2 + \vert \sigma \vert (p^{+}(\sigma) 2n + p^{-}(\sigma)2(n+1)) }\; .
\eea
Note that the set of eigenvalues is the same for either sign of $\sigma$: they are just offset by one increment of $n$. 
Moreover there is a degeneracy, since in there is no dependence of the eigenvalues on $l$. 
The associated eigen-spinors read
\bea
\psi_{k,l,n}=C
\begin{pmatrix}
u_{k,l,n} \\
p^+(\sigma) \alpha  u_{k,l,n-1} + p^-(\sigma) \beta  u_{k,l,n+1} 
\end{pmatrix}, 
\eea
where  $C\in\mathbb{C}$ is a normalization constant and, 
\bea
\alpha =-& \frac{ \frac{\sigma}{\f} + \sqrt{ \left( \frac{\sigma}{f} \right)^2 + \vert \sigma \vert 2n }}{ \vert \sigma \vert^{\frac{1}{2}} \sqrt{2n}} \\
\beta =-& \frac{ \frac{\sigma}{\f} + \sqrt{ \left( \frac{\sigma}{f} \right)^2 + \vert \sigma \vert 2(n+1) }}{ \vert \sigma \vert^{\frac{1}{2}} \sqrt{2(n+1)}}\; .
\eea
%

\subsection{Non-trivial metric}\label{sec:2.3}

For the most general metric given in \eqref{expl}, the Dirac operator takes the form
\eq{ \label{dc1}
D = \gamma^a E^{m}{}_a (\partial_{m} + \tfrac{1}{4}\omega_{m b c}\gamma^{bc} )
\; ,}
where  $E^{m}{}_a\equiv (E^{-1})^{m}{}_a$ is the inverse vielbein  associated with \eqref{expl}, so that \cite{Andriot:2016rdd}, 
\eq{E^{-T}\del = \left(\begin{array}{c}V_1 \\  V_2 \\ V_3- aV_1 - b V_2   \end{array}\right)\; ,  \label{vectors}}
with the $V_a$'s given in \eqref{explv}. Moreover the independent non-vanishing components of the flat spin connection $\omega_{a b c}=E^{m}{}_a \omega_{m b c}$ read
\eq{\spl{\label{conn}
\omega_{112} &= -\f a~;~~~     
\omega_{113} = \f ab~;~~~ 
\omega_{123} = \tfrac{1}{2} \f (-a^2 + b^2 +1)\\
\omega_{212} &= -\f b~;~~~ 
\omega_{213} = -\tfrac{1}{2} \f (a^2 - b^2 +1)~;~~~ 
\omega_{223} =  -\f ab\\
\omega_{312} &= \tfrac{1}{2} \f (a^2 + b^2 -1)~;~~~ 
\omega_{313} = \f b~;~~~
\omega_{323} = -\f a
\; .}}
Taking \eqref{vectors}, \eqref{conn} into account, \eqref{dc1} reduces to
\eq{ \label{dc2}
D=\gamma^1 V_1 + \gamma^2 V_2   + \gamma^3 \big(V_3- aV_1 - b V_2 \big) + i\frac{\f}{4} (1 +a^2 +b^2)
\; .}
As in \S \ref{sec:2.2}, the spectrum falls into two distinct series, depending on whether or not there is non-trivial dependence on the $x^3$ coordinate. 
Let us first examine  the $x^3$-independent case. We expand the Dirac spinor as follows 
\eq{\label{ds246}
\psi_{p,q}=
\begin{pmatrix}
\alpha \\
\beta 
\end{pmatrix} V_{p,q}\; ,
}
for some  constants $\alpha$, $\beta\in\mathbb{C}$. 
The action of the $V_a$ on the $V_{p,q}$-basis \eqref{52} reads 
\eq{
V_1V_{p,q}=iP V_{p,q} ~;~~~ V_2V_{p,q}=iQ V_{p,q} ~, ~~~V_3V_{p,q}=0\; ,
} 
so that the eigenvalue equation
\eq{\label{r4} D \psi=\lambda_{p,q}\psi \; , }
reduces to
\eq{\spl{
-aP\alpha+(P -i Q) \beta +\big[\i\lambda_{p,q} + \tfrac{1}{4}\f(a^2+b^2+1)\big] \alpha &= 0  \\
bQ\beta+(P +i Q) \alpha +\big[\i\lambda_{p,q} + \tfrac{1}{4}\f(a^2+b^2+1)\big] \beta &= 0
\; ,}}
where we have taken \eqref{dc2} into account. From this system we deduce the eigenvalues 
\boxedeq{\label{2.38}
 -\i \lambda_{p,q}=  \tfrac{1}{4}\f(a^2+b^2+1)+ \tfrac{1}{2}(bQ-aP)\pm \sqrt{P^2+Q^2+\tfrac14(aP+bQ)^2}\; .
}
The Dirac eigen-spinor \eqref{ds246} is  determined,  up to an overall normalization constant, by the equation 
\eq{
\beta=\frac{ \tfrac{1}{2}(aP+bQ)\pm  \sqrt{P^2+Q^2+\tfrac14(aP+bQ)^2}}{P-\i Q}\alpha
\; .}
In the case of nontrivial $x^3$-dependence the relevant basis is given by the polynomials \eqref{uultcap2}.  In order not to clutter the notation, we will present the calculation of the spectrum for the case $\sigma>0$, cf.~\eqref{lambda}. The $U$-polynomials obey
\eq{\spl{\label{2.31}
(V_3-aV_1-bV_2) U_{k,l,n} &= \kappa  U_{k,l,n} +\sqrt{\tfrac{ \sigma }{2}} \left(w^*\sqrt{(n+1)} ~\!U_{k,l,n+1} -w \sqrt{n} ~\!U_{k,l,n-1} \right) \\
(V_1+ iV_2) U_{k,l,n} &= \kappa  zU_{k,l,n} +i\sqrt{\tfrac{ \sigma }{2}} \left( \sqrt{(n+1)}(B_-+bw^*)U_{k,l,n+1} + \sqrt{n} (B_+-bw)U_{k,l,n-1} \right) \\
(V_1- iV_2) U_{k,l,n} &= \kappa  z^*U_{k,l,n} +i\sqrt{\tfrac{ \sigma }{2}} \left( \sqrt{(n+1)}(B_+-bw^*)U_{k,l,n+1} + \sqrt{n} (B_- +bw)U_{k,l,n-1} \right)
\; ,}}
where we have defined
\eq{
z:= a+ib~;~~~
\kappa:= \frac{2\pi iK}{1+|z|^2}~;~~~w:=b+\frac{ia}{\sqrt{1+|z|^2}}~;~~~B_\pm:=\pm(1+b^2)+\frac{1+b^2}{\sqrt{1+|z|^2}}
\; .}
Let us now come to the Dirac eigenvalue problem. We start with the following ansatz for the Dirac spinor:
\bea
\psi=
\begin{pmatrix}
\alpha  U_{k,l,n} +\beta  U_{k,l,n-1}\\
\gamma  U_{k,l,n-1} +\delta  U_{k,l,n}
\end{pmatrix}\; .
\eea
Imposing
\bea
\big[D-i\frac{\f}{4}(1+|z|^2)\big]\psi=
\begin{pmatrix}
\alpha'  U_{k,l,n} +\beta'  U_{k,l,n-1}\\
\gamma'  U_{k,l,n-1} +\delta'  U_{k,l,n}
\end{pmatrix}\; ,
\eea
for arbitrary coefficients $\alpha'$, $\beta'$, 
$\gamma' $, $\delta' $, taking \eqref{2.31} into account, 
leads to a system of four homogeneous equations for the four coefficients $\alpha$, \dots, $\gamma$. 
Noting the identity
\eq{
(B_-+bw)(B_+-bw)=w^2\; ,
}
this system turns out to be equivalent to the following two conditions
\eq{\label{2.44}
\beta=\frac{i(B_-+bw)}{w}~\!\gamma~;~~~
\delta=\frac{i(B_-+bw^*)}{w^*}~\!\alpha
\; ,}
where it is  assumed that $a, b$ are not both zero. Imposing in addition the eigenvalue equation
\eq{\label{r3} D \psi=\lambda_{k,l,n}\psi~, }
taking \eqref{2.44} into account, leads to a system of four homogeneous equations for $\alpha, \gamma$. Clearly this is highly overdetermined. Remarkably, however, the system admits a nontrivial solution, provided
\boxedeq{
-i\lambda_{k,l,n} =\frac{1}{4}\f (1+|z|^2) \pm \sqrt{ \frac{4\pi K^2}{1+|z|^2} +4\pi n K\f\sqrt{1+|z|^2} }
\; ,}
where we took \eqref{lambda} into account. 
As was observed in the case of the scalar spectrum \cite{Andriot:2016rdd}, the eigenvalues depend on the $a, b$ parameters only through the norm of $z$.~As in \S \ref{sec:2.2} we see that there is a degeneracy, since the eigenvalues are independent of $l$. 

\section{Reduction to four dimensions}\label{sec:3}

We would now like to examine the 4D  effective theory arising as the low-energy limit of a 7D gauge-fermion theory compactified on the Heisenberg nilmanifold $\mathcal{M}_3$. The 7D Lagrangian consists of a Yang-Mills term $\mathcal{L}^{\text{YM}}_{7\text{D}}$ 
and a fermion term $\mathcal{L}^\text{f}_{7\text{D}}$. 
The effective theory in four dimensions,  $\mathcal{L}^{\text{eff}}_{4D}$, will be given by 
\eq{\label{32}
\mathcal{L}_{4D}^{\text{eff}}= 
\int\d^3y~\big(
\mathcal{L}^{\text{YM}}_{7\text{D}}+
\mathcal{L}^{\text{f}}_{7\text{D}}
\big)
\; .
}
The  
right-hand side above indicates the KK reduction of the  seven-dimensional theory, and involves integrating over the three-dimensional internal space 
parameterized by the $y$-coordinates. 

Moreover, we will place ourselves in the  small fiber/large base limit \cite{Andriot:2018tmb}, 
\eq{
\label{limit}
|\f| \ll \frac{1}{r^i}~, ~i=1,2,3 \quad \Rightarrow \quad \frac{1}{r^{1,2}} \ll \frac{1}{r^3}
\; .}
In this limit all fields whose masses carry an $r^i$ dependence (i.e.~all the KK modes) decouple, leaving in the theory only those fields with masses of either the order of $|\f|$, or zero. Explicitly, the reduction ansatz for the gauge fields is given by \cite{Andriot:2020ola}:
\eq{
\mathcal{A}^a=\frac{1}{\sqrt{V}}\Big(
A^a+\sum_{I=1}^3\phi^{a I}\tilde{E}^I
\Big)
\; ,}
where $A^a$, $a=1,\dots\text{dim}(G)$, is a 4D one-form and $\phi^{a I}$, $I=1,2,3$, are three scalars in the adjoint of the Lie algebra of $G$. 
The $\tilde{E}^I$'s span the space of low-lying one-forms on $\mmm_3$, cf.~\ref{sec:b}. They can be chosen so that 
$\tilde{E}^{1,2}$ are harmonic and $\tilde{E}^3$ has Laplacian eigenvalue\footnote{In the present paper we reinstate 
dependence on the the $a$, $b$ parameters which were set to zero in \cite{Andriot:2020ola}.} $f^2(a^2+b^2+1)^2$. 
Integrating the Yang-Mills term over the internal space $\mathcal{M}_3$ we obtain \cite{Andriot:2020ola}:
\eq{\int\d y^3 \mathcal{L}^{\text{YM}}_{7\text{D}}=
\tfrac{1}{2} F^a_{\mu\nu}F^{a\mu\nu}+\sum_{I=1}^3D_\mu\phi^{aI} D^\mu\phi^{aI}+M^2(\phi^{a3})^2+\mathcal{U}
 \; ,}
where,
\eq{
 \mathcal{U}=\text{Tr}\Big( -2i g M[\phi^1,\phi^2]\phi^3 + \tfrac12 g^2\sum_{I,J=1}^3[\phi^I,\phi^J][\phi^I,\phi^J]\Big)
\; ,}
with $F^a_{\mu \nu} = 2 \partial_{ [ \mu} A^a_{\nu ]} +ig f^a{}_{bc} A^b_{\mu} A^c_{\nu}$,  
 $D_{\mu} \phi^{aI}= \partial_{\mu} \phi^{aI}+i g f^a{}_{bc}A^b_{\mu} \phi^{cI}$, and $M= \vert \f \vert (a^2+b^2+1)$;  $f^a{}_{bc}$ are the structure constants of the algebra of the gauge group. 
The 4D gauge coupling constant $g$ is related to the 7D coupling and the volume of $\mmm_3$ via
\eq{\label{4dcoupling}
g=\frac{ g_{7\text{D}} }{\sqrt{V}}
\; .}
The fermionic Lagrangian will be taken to be of the form
\eq{\label{ferm1}
\mathcal{L}^{\text{f}}_{7\text{D}}=\overline{\psi}_i \Gamma^M (\delta^{ij}\nabla_M +i\mathcal{A}_M^a\rho_a^{ij})\psi_j
 +\tfrac16 F_{MNP}\overline{\psi}_i\Gamma^{MNP}\psi_i + M_{0}\overline{\psi}_i \psi_i
\; ,}
where the hermitian matrices $\rho_a^{ij}$ provide a representation $R$ of the {Lie algebra} of the gauge group, so that 
$i,j=1,\dots,\text{dim}(R)$, and $\psi_i$ transforms in the $R$ representation. 
We have also allowed for  a constant background  zero-form flux (a mass term) $M_0$, and a three-form flux $F_{MNP}$, which will be assumed to be along the internal manifold $\mathcal{M}_3$ in order not to break the 4D Lorentz invariance.\footnote{We may also allow for 
other types of fluxes, however this will result in a similar mass terms  as in \eqref{132} below, so we do not  introduce them independently.} This implies that the only non-vanishing component is given by
\eq{F_{mnp} =  M_1 \varepsilon_{mnp}\; , }
for some real constant $M_1$, and the Levi-Civita symbol is a tensor in our conventions. 

Our reduction ansatz for  the 7D spinors is as in \eqref{a20},
\eq{
\psi_i=(\chi_{i+}  +\theta_{i-} )\otimes \xi \; ,
}
where $\chi_i$, $\theta_i$ are Weyl 4D spinors and $\xi$ is a   spinor on $\mathcal{M}_3$.  
We can already see that the resulting 4D model will  necessarily be non-chiral since, the 4D positive and negative chiralities of the spinors both transform in the same representation $R$.

We expand $\xi$ on a basis of normalized eigen-spinors of the Dirac operator in 3D
\eq{
\sigma^m\nabla_m\xi=\lambda \xi
\; ,}
where the eigenvalues of the Dirac operator on $\mathcal{M}_3$ were given in \S\ref{sec:2.3}. In the limit \eqref{limit} of decoupling of the KK modes, only 
the lowest eigenspinor is kept, corresponding to eigenvalue $\lambda=i\frac{\f}{4} (a^2+b^2 +1)$, cf.~\eqref{2.38}. Moreover we assume that 
$\xi$ is normalized: $\int\!\d^3y~\!\xi^\dagger\xi=1$. 
Putting everything together we obtain
\eq{
  \int\d^3y~\mathcal{L}^{\text{f}}_{7\text{D}}=\mathcal{L}_{\text{kinetic}}+\mathcal{L}_{\text{Yukawa}}+\mathcal{L}_{\text{mass}}
\; ,}
where
\eq{\spl{
\mathcal{L}_{\text{kinetic}}&=\overline{\chi}_{i+} \gamma^\mu (\delta^{ij}\partial_\mu +i g {A}_\mu^a\rho_a^{ij}){\chi}_{j+}
+\overline{\theta}_{i-} \gamma^\mu (\delta^{ij}\partial_\mu + i g{A}_\mu^a\rho_a^{ij}){\theta}_{j-} 
\\
\mathcal{L}_{\text{Yukawa}}&=ig(\overline{\theta}_{i-}\chi_{j+} -\overline{\chi}_{i+}\theta_{j-} )\rho_a^{ij}\Phi^a
\\
\mathcal{L}_{\text{mass}}&=M_c~\overline{\theta}_{i-}\chi_{i+}  +M_c^*~ \overline{\chi}_{i+}\theta_{i-} \; ,}}
and we have defined a complex ``mass'' $M_c$  
\eq{\label{4dcoupling}
 M_c=M_0+i\big[M_1+\tfrac{\f}{4} (a^2+b^2 +1)\big]
\; .}
The adjoint scalar $\Phi^a$ is defined by
\eq{\label{defphi}
\Phi^a:=\sum_{I=1}^3\phi^{a I}c^I \; ,
}
where the three constants $c^I$ are given by $c^I:=\xi^{\dagger}\sigma^I\xi$. 
To make this more explicit, let us give a parameterization for $\xi$. Up to an unimportant overall phase we may set 
\eq{
\xi=\begin{pmatrix}
e^{-i\beta}\cos\frac{\alpha}{2} \\
\sin\frac{\alpha}{2}
\end{pmatrix}\; ,
}
for some angles $\alpha$, $\beta$. We then find $c^1=\sin\alpha\cos\beta$, $c^2=\sin\alpha\sin\beta$, $c^3=\cos\alpha$. I.e.~$\vec{c}$ can be 
thought of as a unit vector of $\mathbb{R}^3$.  

The final expression for the fermionic 4D theory can also be expressed in terms of 4D Dirac spinors $\Psi_i:=\i\chi_{i+}  +\theta_{i-} $ 
\eq{\label{132}
  \int\d^3y~\mathcal{L}^{\text{f}}_{7\text{D}}=\overline{\Psi}_i \Gamma^\mu (\delta^{ij}\partial_\mu +i g{A}_\mu^a\rho_a^{ij})\Psi_j
 + g\rho_a^{ij}\Phi^a \overline{\Psi}_i \Psi_j
 +\big[M_1+\tfrac{\f}{4} (a^2+b^2 +1)\big]\overline{\Psi}_i \Psi_i
 -\i M_0 \overline{\Psi}_i \gamma_5\Psi_i
\; .}
The 7D parameters $M_{0,1}$ are free (up to flux quantization), and can be thought of as arising from the inclusion
of constant background flux on the nilmanifold.
 
\begin{appendix}

\section{Fermion conventions}
\label{app:fermions}

In a space of arbitrary dimension and Lorentzian signature, the gamma matrices are taken to satisfy
\eq{
(\Gamma^M)^{\dagger}=\Gamma^0\Gamma^M\Gamma^0\; .
}
We define the antisymmetric product of $n$
gamma matrices by
\eq{
\Gamma_{M_1\dots M_n}:=\Gamma_{[M_1}\dots\Gamma_{M_n]}\; .
}
Given a spinor $\psi$ we define
\eq{
\overline{\psi}:=   \psi^+ \Gamma^0
\; .}
Given a spinor $\psi$ in a space of arbitrary dimension {\it and arbitrary signature}, we define
\eq{
\tilde{\psi}:=   
{\psi}^{\text{Tr}}C \; ,}
where $C$ is the charge conjugation matrix. This has the property that for any spinors $\psi$, $\chi$, the bilinear $\tilde{\psi}\Gamma_{M_1\dots M_n}\chi$ is an antisymmetric tensor of order $n$.

\subsection{Spinors in 4D Minkowski space}

The charge conjugation matrix in $1+3$ dimensions satisfies
\eq{\label{a5}
C^{\text{Tr}}=-C; ~~~~~~ (C\gamma^\mu)^{\text{Tr}}=-C\gamma^\mu\; .
}
The chirality matrix is defined by
\eq{
\gamma^5:=i\gamma^0\dots\gamma^3; ~~~~~~(\gamma^5)^2=1 \; .
}
The fundamental, positive-chirality (Weyl), two-component, spinor representation 
$\psi_+$ is complex, meaning that its complex conjugate $\psi_-$  has negative chirality.
The complex conjugate $\psi_-$  of $\psi_+$  is {\it defined} by
\eq{\label{scc}
\tilde{\psi}_-:=
\overline{\psi}_+ \; ,
}
which also implies 
\eq{
\tilde{\psi}_+=-
\overline{\psi}_- \; .}
We stress that these are {\it not} reality conditions: they simply define $\psi_-$ in terms of $\psi_+$ or vice-versa. 
Indeed a reality condition would equate (up to a constant) $\overline{\psi}_+$ and $\tilde{\psi}_+$, which is impossible in four dimensions. 

Let $\chi_\pm$, $\psi_\pm$ be arbitrary {\it anticommuting} Weyl spinors of positive or negative chirality. We have the following useful relations 
\eq{
\tilde{\psi}_\pm\chi_\mp=0~;~~~\tilde{\psi}_\pm\gamma_\mu\chi_\pm=0
\; .}
The following symmetry relations are valid for Weyl spinors of any chirality
\eq{
\tilde{\psi}\chi =\tilde{\chi}\psi ~;~~~\tilde{\psi}\gamma_\mu\chi=\tilde{\chi}\gamma_\mu\psi
\; .}
It is also useful to note the following complex conjugation relations
\eq{\label{ccra}
(\overline{\psi}_\pm\gamma_\mu\chi_\pm)^*=-\overline{\chi}_\pm\gamma_\mu\psi_\pm
~;~~~
(\overline{\psi}_\pm \chi_\mp)^*=\overline{\chi}_\mp \psi_\pm
\; .}
Let us now consider an arbitrary anticommuting  Dirac spinor $\psi_D$. It can be written in terms of two arbitrary 
Weyl spinors $\chi_+$ and $\theta_+$
\eq{\label{a12}
\psi_D=\chi_++\theta_- \; ,}
where  $\theta_-$ is  the complex conjugate of $\theta_+$, given by \eqref{scc}. The ``Dirac mass'' is given by
\eq{\label{dirm}
\overline{\psi}_D\psi_D=\overline{\chi}_+\theta_-  +  \overline{\theta}_-\chi_+
=\tilde{\chi}_-\theta_-  -  \tilde{\theta}_+\chi_+
\; ,}
which is real, as can be verified using \eqref{ccra}.

A Weyl spinor can be considered as a special case of a Dirac spinor whose component of negative or positive chirality vanishes. 
Therefore it is sometimes said that the Dirac mass of a Weyl spinor vanishes: indeed setting $\chi$ or $\theta$ to zero would make 
the right hand side of \eqref{dirm} vanish. Nevertheless a mass term can be defined for a single Weyl spinor: it suffices to set $\theta_\pm=\chi_\pm$ in \eqref{dirm}. This is  sometimes described as defining a Majorana spinor
\eq{\psi_M=\chi_++\chi_-\; ,}
which is nothing other than a Dirac spinor whose negative-chirality component is the complex conjugate of its positive-chirality component.\footnote{In our conventions the Majorana spinor satisfies the reality condition: 
$\overline{(\gamma_5\psi_M)}=\tilde{\psi}_M$. 
%
}
A real mass term for a Weyl spinor $\chi$ can then be written  in terms of  $\psi_M$,
\eq{
\overline{\psi}_M\psi_M=\overline{\chi}_+\chi_-  +  \overline{\chi}_-\chi_+
=\tilde{\chi}_-\chi_-  -  \tilde{\chi}_+\chi_+
\; .}
This is sometimes called the Majorana mass.

\subsection{Spinors in 3D Riemannian space}

In a 3D space of Euclidean signature the gamma matrices can be taken to be the Pauli matrices, 
while the charge conjugation matrix can be taken as $C=i\sigma_2$. We have
\eq{
C^{\text{Tr}}=-C; ~~~~~~ (C\gamma^m)^{\text{Tr}}=C\gamma^m \; .
}
The irreducible spinor representation of $Spin(3)$ has two {\it complex} components. 
We thus have the following useful symmetry properties
\eq{
\tilde{\psi}\gamma^{m_1\dots m_p}\chi=(-1)^{\frac12 (p-1)(p-2)}\tilde{\chi}\gamma^{m_1\dots m_p}\psi
\; ,
}
where $\chi$, $\psi$ are arbitrary {\it commuting} spinors. We define the complex conjugate $\psi_c$ of $\psi$ via
\eq{\tilde{\psi}_c
=\psi^\dagger\; ,}
so that $\psi_c$ transforms as a spinor. 
We then have the complex conjugation properties,
\eq{\spl{
(\tilde{\psi}_c\gamma^{m_1\dots m_p}\chi)^*&=-(-1)^{p}\tilde{\psi}\gamma^{m_1\dots m_p}\chi_c
\; .
}}

\subsection{Spinors in 7D Lorentzian space}

The irreducible spinor representation of $Spin(1,6)$  has eight complex components. In terms of an $Spin(1,6)\rightarrow Spin(1,3)\times Spin(3)$ decomposition, the 7D spinor $\psi$ decomposes as
\eq{\label{a20}
\psi=(\chi_+  +\theta_- )\otimes \xi \; ,
}
where $\chi$, $\theta$ are irreducible Weyl spinors of $Spin(1,3)$ and $\xi$ is an irreducible spinor of $Spin(3)$.  
The seven-dimensional gamma matrices $\Gamma^M$ decompose as
\eq{
\Gamma^\mu=\gamma^\mu\otimes\mathbb{I}_2~;~~~
\Gamma^{m+3}=\gamma^5\otimes\gamma^m
\; ,}
where $\mu=0,\dots, 3$ and $m=1,2,3$.

\section{Laplacian eigen one-forms}\label{sec:b}

In this section we work out the low-lying Laplacian eigen one-forms  (i.e.~those which do not descend form KK states) in the case of non-trivial 
metric. These are   linear combinations of the coframe associated with the metric \eqref{expl}
\eq{\label{b1}
A=\sum_{a=1}^3c_aE^a
\; ,}
where the $c_{a}$'s  are real constants to be determined in the following and, 
\eq{\label{b2}
E^1=e^1+a e^3~;~~~E^2=e^2+b e^3~;~~~E^3=e^3\; .
}
Recall that in the  case of an undeformed metric ($a$,$b=0$) the low-lying Laplacian  eigen one-forms are $e^{1,2}$, which are harmonic, and $e^3$, which has 
eigenvalue $\f^2$ \cite{Andriot:2018tmb}. To see how this spectrum is modified for a general metric ($a$,$b\neq0$), we need to calculate the action of the Laplacian $\Delta$ on $A$ 
\eq{
\Delta A=\sum_{a=1}^3c_a(\d\d^{\dagger}+\d^{\dagger}\d)E^a\; ,
}
 where $\d^{\dagger}\equiv\star\d\star$. The Hodge star is calculated with respect to the deformed metric \eqref{expl}, and operates canonically on $E^a$
 \eq{
 \star E^a=\frac12 \sum_{b,c=1}^3\varepsilon^{abc}E^b\wedge E^c\; ,
 }
while the action of the exterior differential on $E^a$ is  calculated form \eqref{mcheis}, \eqref{b2}. It is then easily verified that 
the coframe is co-closed, $\d^{\dagger}E^a=0$, and,
\eq{
\Delta A=\f^2(a^2+b^2+1)(ac_1+bc_2+c_3)(aE^1+bE^2+E^3)\; .
}
Setting $c_1=a$, $c_2=b$, $c_3=1$, it follows that $aE^1+bE^2+E^3$ is a Laplacian eigen one-form with eigenvalue $\f^2(a^2+b^2+1)^2$. Moreover,  we obtain a two-dimensional space of harmonic one-forms, parameterized by the solutions of $ac_1+bc_2+c_3=0$. 

\end{appendix}

\bibliography{refs}
\bibliographystyle{unsrt}

\end{document}